УДК: 51.76, 53.096

# Условия существования поляронных состояний в классических молекулярных цепочках при конечных температурах

**Лахно В.Д.**[*], **Фиалко Н.С.**

*Федеральное государственное бюджетное учреждение науки Институт математических проблем биологии Российской академии наук*

***Аннотация.*** В современной литературе поляронные состояния в молекулярных цепочках в подавляющем большинстве рассчитываются для нулевой температуры. При этом считается, что их свойства мало меняются, если температура отлична от нуля, но много меньше характерной энергии, равной глубине поляронного уровня. Однако результаты вычислительных экспериментов приводят к предположению, что в неограниченно длинной цепочке поляронные состояния разрушаются при сколь угодно мало отличающейся от нуля температуре. Данная статья посвящена разрешению описанной «парадоксальной» ситуации.

***Ключевые слова:*** *поляронное состояние, статистическая сумма, гамильтониан Холстейна, уравнение Ланжевена.*

В работах [1–3] было проведено детальное исследование динамики формирования поляронных состояний в молекулярных цепочках при температуре, равной нулю. В работе [4] были предприняты попытки обобщить результаты [1–3] на случай конечных температур. С этой целью для описания динамики молекулярной цепочки использовались уравнения Ланжевена. Результаты вычислительных экспериментов привели к странным результатам.

Если сказать коротко, то они свидетельствовали о том, что в неограниченно длинной цепочке поляроннные состояния разрушаются при сколь угодно мало отличающейся от нуля температуре. Это полностью противоречит бытующему представлению о том, что поляронные состояния должны разрушаться при температуре, по порядку величины равной выигрышу в энергии, который получает электрон, образуя поляронное состояние.

Данная статья посвящена разрешению описанной «парадоксальной» ситуации.

Исходной для моделирования динамики поляронных состояний в молекулярной цепочке была полуклассическая модель Холстейна с гамильтонианом:

$$\hat{H} = \sum_{n,m} \nu_{nm} |n\rangle\langle m| + \chi \sum_n q_n |n\rangle\langle n| + \frac{K}{2} \sum_n q_n^2, \qquad (1)$$

где $\nu_{nm}$ – матричный элемент перехода между сайтами $n$ и $m$, $\chi$ – константа взаимодействия электрона с колебаниями цепочки, $K$ – упругая постоянная

---
[*] lak@impb.psn.ru

В приближении ближайших соседей спектр гамильтониана (1) хорошо известен и имеет вид:

$$E = (E_0, \lambda_k), \qquad \lambda_k = -2\nu \cos\frac{k\pi}{N+1}, \qquad k = 1,\ldots N, \qquad (2)$$

где $E_0$ – поляронный уровень энергии, $\lambda_k$ – спектр делокализованных состояний в недеформированной цепочке.

Вычислим статистическую сумму для спектра (2):

$$Z = \sum_n \exp\left(-\frac{E_n}{k_\text{B}T}\right) = \exp\left(-\frac{E_0}{k_\text{B}T}\right) + Z_1, \qquad Z_1 = \sum_{k=1}^{N} \exp\left(-\frac{\lambda_k}{k_\text{B}T}\right), \qquad (3)$$

где $T$ – температура. Подстановка (2) в (3) дает для величины $Z_1$ следующее приближенное выражение:

$$Z_1 = (N+1) J_0\left(i\frac{2\nu}{k_\text{B}T}\right), \qquad (4)$$

где $i$ – мнимая единица. Выражение (4) является точным для достаточно длинных цепочек: $N \gg 1$. В интересующем нас случае достатчно низких температур ($2\nu/k_\text{B}T \gg 1$) для функции Бесселя можно воспользоваться асимптотическим выражением:

$$J_0\left(i\frac{2\nu}{T}\right) \approx \frac{\exp(2\nu/k_\text{B}T)}{\sqrt{4\pi\nu/k_\text{B}T}}. \qquad (5)$$

Из (3)–(5) следует, что заселенность поляронного уровня $P_0 = Z^{-1}\exp(-E_0/k_\text{B}T)$ равна:

$$P_0 = \left(1 + \frac{N+1}{\sqrt{4\pi\nu/k_\text{B}T}} \exp\left(\frac{E_0 + 2\nu}{k_\text{B}T}\right)\right)^{-1}. \qquad (6)$$

Из (6) следует вывод, который без проведенного простого анализа мог показаться удивительным: в бесконечной цепочке поляронное состояние не существует при сколь угодно малой температуре.

В современной литературе поляронные состояния в $D$-мерных цепочках в подавляющем большинстве рассчитываются для $T = 0$. При этом считается, что их свойства мало меняются, если температура отлична от нуля, но много меньше характерной энергии, равной глубине поляронного уровня $E_0$.

С математической точки зрения речь идет о том, что если рассматривать вероятность $P_0$ как функцию двух переменных: $P_0 = P_0(N, T)$, то предельный переход к точке $P_0 = P_0(\infty, 0)$ будет разным в зависимости от того, в каком порядке мы устремляем $N$ и $T$ к своим предельным значениям. А именно:

$$\lim_{N\to\infty} \lim_{T\to 0} P_0(N,T) = 1,$$

$$\lim_{T\to 0} \lim_{N\to\infty} P_0(N,T) = 0.$$

В работе [4] моделировался развал полярона для polyG/polyC-цепочек ДНК. Оценим критическое значение длины цепочки $N_c$ для polyG/polyC при $T = 300\,\text{K}$. Согласно [5], в этом случае: $\nu = 0.084$ эВ, $E_0 + 2\nu = -1.7 \cdot 10^{-2}$ эВ, $k_\text{B}T \simeq 0.0259$ эВ. В результате для $N_c$ получим:

$$N_c = \sqrt{\frac{4\pi\nu}{k_\text{B}T}} \exp\left(-\frac{E_0 + 2\nu}{k_\text{B}T}\right) \approx 12.$$

Условие существования поляронных состояний в этом случае приводит к очень коротким цепочкам с длиной $N \ll 12$.